\documentclass[aps,twocolumn,superscriptaddress,longbibliography,amsmath,amssymb,amsfonts,citeautoscript]{revtex4-2}
\usepackage{graphicx}
\usepackage{bm}
\usepackage{bbm}
\usepackage{color}
\usepackage{epstopdf}
\usepackage{amsmath}
\usepackage{amssymb}
\usepackage{ulem}
\usepackage[urlcolor=blue,colorlinks=true,citecolor=blue,linkcolor=blue,pdfstartview={FitH},bookmarks=false]{hyperref}
\usepackage{todonotes}
\usepackage{ulem}
\newcommand{\GFzub}[2]{\langle\!\langle #1;#2\rangle\!\rangle}
\newcommand{\be}{\begin{equation}}
\newcommand{\ee}{\end{equation}}
\newcommand{\bea}{\begin{eqnarray}}
\newcommand{\eea}{\end{eqnarray}}

\newcommand{\fig}[1]{Fig.~\ref{#1}}
\newcommand{\figs}[1]{Figs.~\ref{#1}}
\newcommand{\en}{\varepsilon}
\newcommand{\w}{\omega}
\newcommand{\s}{\sigma}
\newcommand{\G}{\Gamma}
\newcommand{\up}{\uparrow}
\newcommand{\down}{\downarrow}

\begin{document}

\title{Nonlocal correlations transmitted between quantum dots\\ via short topological superconductor}

\author{G. G\'{o}rski}
\affiliation{Institute of Physics, College of Natural Sciences, University of Rzesz\'{o}w, 35-310 Rzesz\'{o}w, Poland}

\author{K.P. W{\'o}jcik}
\affiliation{Institute of Molecular Physics, Polish Academy of Sciences,  60-179 Pozna{\'n}, Poland}

\author{J.\ Bara\'nski}
\affiliation{Polish Air Force University, ul. Dywizjonu 303 nr 35, 08-521 D\c{e}blin, Poland}

\author{I. Weymann}
\affiliation{Institute of Spintronics and Quantum Information, Faculty of Physics, A.~Mickiewicz University, 61-614 Pozna{\'n}, Poland}

\author{T. Doma\'nski}
\affiliation{Institute of Physics, M. Curie-Sk\l{}odowska University, 20-031 Lublin, Poland}

\date{\today}

\begin{abstract}
We study the quasiparticle states and nonlocal correlations of a hybrid structure, comprising two quantum
dots interconnected through a short-length topological superconducting nanowire hosting 
overlaping Majorana modes. We show that the hybridization between different components of this setup gives rise to the emergence of molecular states, which are responsible for nonlocal correlations. We inspect the resulting energy structure, focusing on the inter-dependence between the quasiparticles of individual quantum dots. We predict the existence of nonlocal effects, which could be accessed and probed by crossed Andreev reflection spectroscopy.
Our study would be relevant 
to a recent experimental realization of the minimal Kitaev model 
[T. Dvir {\it et al.}, \href{https://doi.org/10.1038/s41586-022-05585-1}{Nature 
{\bf 614}, 445 (2023)}], by considering its hybrid structure with side-attached 
quantum dots.
\end{abstract}


\flushbottom
\maketitle


Majorana quasiparticles\cite{Majorana1937Apr,Wilczek2009Sep}, emerging at the boundaries of topological superconductors, 
are currently a topic of intensive studies, motivated by perspectives of using them as 
stable quantum bits (immune to decoherence due to topological protection) 
and in quantum computing (owing to their non-Abelian character). \cite{Kitaev2003Jan}
Experimental signatures of the zero-energy Majorana modes have been reported in numerous systems, ranging from (i) semiconducting nanowires and/or nanochains of magnetic atoms contacted with conventional superconductors, (ii) interfaces of the planar Josephson junctions or (iii) outer boundaries of magnetic islands deposited on superconducting surfaces [see Refs.~\cite{Flensberg_2021,Kouwenhoven-2020,Lutchyn2018May,Aguado_2017} for a comprehensive overview].
Another route for obtaining the Majorana modes is possible
in vortices of triplet superconductors. Theoretical and experimental advancements in this direction include e.g.\ topological insulator/s-wave superconductor
heterostructures \cite{FuKane_2008,Sun_etal_2016} and intrinsic superconducting topological insulators, such as FeTeSe 
\cite{Wang_etal_2018,Shiyu_etal_2020,Kong_etal_2021,Xu_etal_2016,Hu_etal_2022}.
Majorana modes have been also predicted to emerge in more complex magnetic textures proximitized to superconducting materials, such as those of the skyrmion geometry.
In all of these platforms the Majorana quasiparticles do always appear in pairs. It has not been firmly established, however, whether they are mutually correlated over certain spatial or temporal scales. Some consequences caused by a hybridization of the Majorana modes confined in nearby vortices have been so far discussed in Refs \cite{DasSarma_2009,Hu_etal_2016,Fu_etal_2023}, inspecting extent of their coherence.

A convenient (although indirect) method for probing the interdependence of the Majorana boundary modes is to exploit 
hybrid nanostructures, such as those composed
of quantum dots side-coupled to topological superconductors. Leakage of the Majorana modes into these 
objects has been predicted theoretically \cite{Vernek-2014} and confirmed experimentally \cite{Deng-2016}. 
Such approach has stimulated further experimental studies \cite{Deng-2018}, inspecting the nonlocality of emergent Majorana
modes hybridized with quantum dot states. 
From the characteristic behavior of their energy spectra,
associated with splittings and anticrossings,
it is possible to extract the degree of nonlocality of Majorana quasiparticles quantified by the spin canting angle, as discussed in the next section.
Such canting can be manifested either by a partially separated Andreev bound state or by a nonlocal state consistent with the well-developed Majorana mode.

\begin{figure}
\centering
\includegraphics[width=0.7\linewidth]{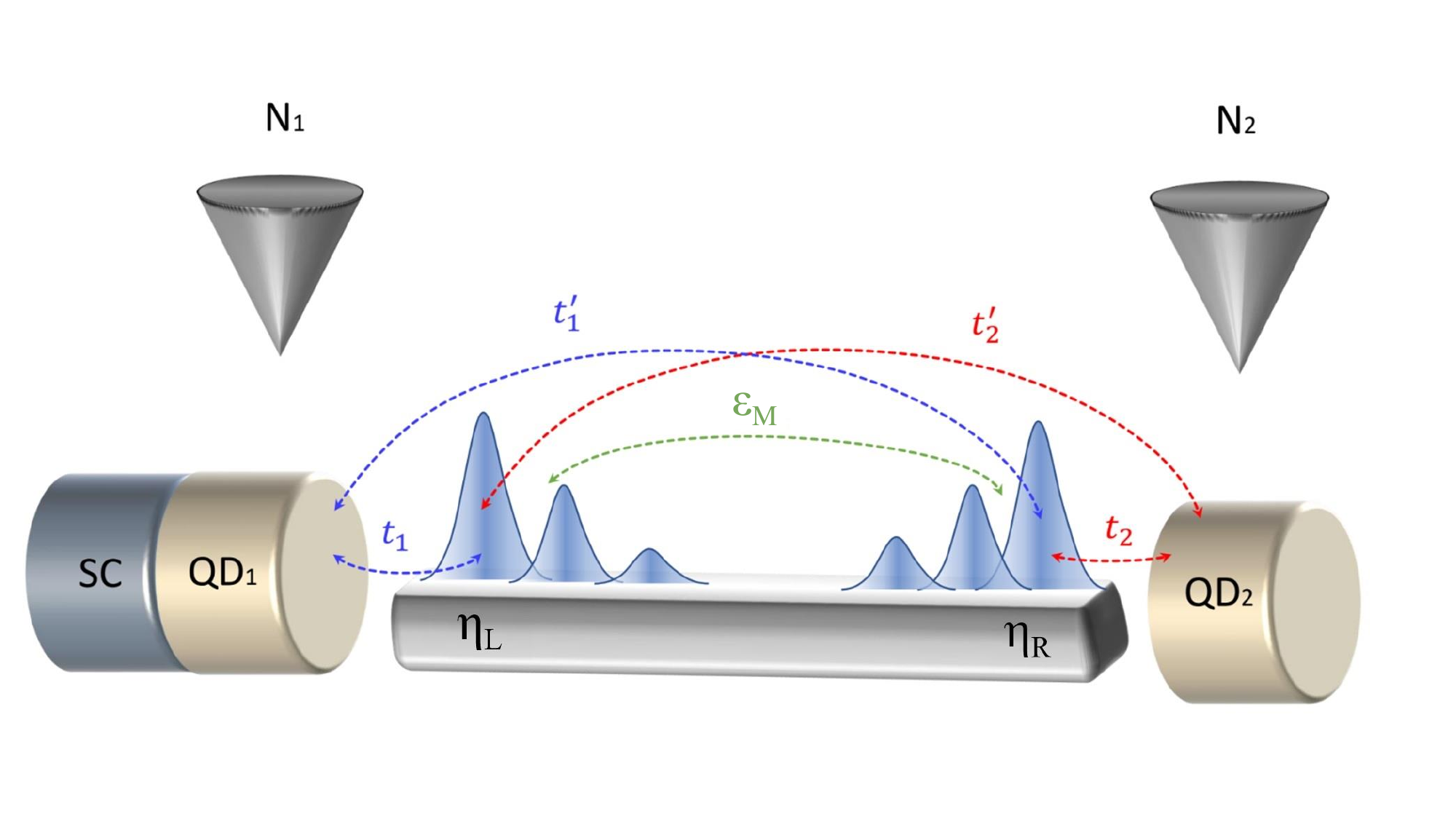}
\caption{Schematic of the considered hybrid nanostructure, consisting of two quantum dots (QD$_{1,2}$) attached to the opposite sides of the superconducting topological nanowire, hosting Majorana modes described by the operators $\eta_{\rm L}$ and $\eta_{\rm R}$. The electronic spectrum of QD$_{1}$ can be probed by the charge transport measurements imposed by the bias voltage in the N$_{1}$-QD$_{1}$-SC part of this setup.}
\label{scheme}
\end{figure}

To address the problem of nonlocality, here we propose to consider a hybrid nanostructure consisting of two
quantum dots interconnected via the topological superconductor. 
More specifically, 
we focus on the nonlocal correlations transmitted between these quantum dots 
solely through the short-length topological superconductor, allowing for 
finite overlap of the Majorana wave-functions (see Fig.~\ref{scheme}). 
The considered setup would be particularly relevant for 
the minimal-length nanowires, such as realized recently Kitaev chain 
comprising only two or three sites \cite{Kouewenhoven-2023,Bordin2024Feb}.
In such short wires, the Majorana modes indeed do overlap with one another, 
so their leakage onto any side-attached quantum dot(s) would be a source of mutual cross-correlations \cite{Souto_2023}.
Here we identify signatures of the inter-dot correlations suitable for experimental detection via the crossed Andreev spectroscopy.

Our study reveals that the leakage of Majorana modes into the quantum dots induces 
a nonlocal electron pairing, both in the triplet and singlet channels. Furthermore, 
the quasiparticle states of the system are showing up simultaneously 
in both quantum dots, although with different spectral weights, strongly dependent on 
the canting angle. Such nonlocal pairings can induce
crossed Andreev reflections,
empirically indicating distant cross-correlations. 
We determine transmittance of the local
and nonlocal charge transport channels operating
in the subgap region, which provide
information about the conventional and topological 
quasiparticles of this hybrid nanostructure.
We believe that our findings 
shall foster further endeavours 
in the {\it bottom-up} engineering of
topological superconducting nanowires, such as these reported in Refs.~\cite{Kouewenhoven-2023,Bordin2024Feb}.

\section{Model}
\label{sec:model}

The considered hybrid structure, as displayed in Fig.~\ref{scheme},
can be described by the following microscopic Hamiltonian 
\be
\label{eq:H}
H =  H_{\rm MW} + H_{\rm L} + H_{\rm R},
\ee
where $H_{\rm MW}$ models the Majorana wire (MW)
attached to the left ($H_{\rm L}$) and right ($H_{\rm R}$)
quantum dot junctions, respectively.
We assume that the left part is formed by the first quantum dot 
(QD$_{1}$), which is weakly coupled to a normal metallic lead
(N$_1$) and strongly hybridized with conventional ($s$-wave)
superconductor (SC). Under such conditions,
the superconducting proximity effect gives rise to the formation of trivial (finite-energy) bound states at QD$_{1}$  
\cite{Bauer-2007,Yamada-2011,Baranski-2013,Gorski-2018}.
Furthermore, coupling QD$_1$ to the topological superconductor
allows the Majorana mode to leak into this quantum dot,
and by interfering with the trivial bound states,
such subsystem develops joint (`molecular') quasiparticle spectrum. 
We will examine in detail the corresponding
spin-resolved energy structure of QD$_{1}$ junction,
which can be formally described by
\bea \label{eq:HQD1}
H_{\rm L} &=& \sum_{\s} \en_1 d_{1\s}^\dag d_{1\s}
 +\sum_{\mathbf{k}\s}\en_{{\rm N_1}\mathbf{k}} c^\dag_{{\rm N_1}\mathbf{k}\sigma} c_{{\rm N_1}\mathbf{k}\sigma}
\nonumber \\
&+&\sum_{\mathbf{k}\s}\en_{{\rm S}\mathbf{k}} c^\dag_{{\rm S}\mathbf{k}\sigma} c_{{\rm S}\mathbf{k}\sigma}- \Delta_{S}\sum_\textbf{k} ( c^{\dagger}_{{\rm S}\mathbf{k}\uparrow }c^{\dagger}_{{\rm S}-\mathbf{k} \downarrow}+\text{H.c.}) \nonumber \\
&+&\sum_{\mathbf{k}\s} \sum_{\beta={\rm{N_1,S}}} V_{\beta\s} \left(d^\dag_{1 \s}
c_{\beta\mathbf{k}\s} + c^\dag_{\beta\mathbf{k}\s} d_{1 \s} \right).
\eea
Here, the operator $d_{1 \s}$ ($d^\dag_{1 \s}$) annihilates (creates)
an electron with spin $\s$ and energy $\en_1$ at QD$_1$,
the operators $c_{\beta\mathbf{k}\sigma}$ ($c^\dag_{\beta\mathbf{k}\sigma}$) correspond to itinerant electrons with spin $\s$ and momentum 
$\mathbf{k}$ in external electrodes $\beta=\{\rm{N}_1,\rm{S}\}$, while 
$\en_{{\rm \beta}\mathbf{k}}$ denotes the energy of respective electrons and $\Delta_{S}$ stands for the isotropic pairing gap. The last term
in Eq.\ ({\ref{eq:HQD1}) describes the hybridization between the mobile electrons of the leads and QD$_{1}$ with the corresponding tunneling matrix elements $V_{\beta\s}$.

On the other hand, the opposite-side quantum dot (QD$_2$) is 
assumed to be weakly coupled to another normal lead (N$_2$),
and can be modelled by
\bea 
H_{\rm R} &=& \sum_{\s} \en_2 d_{2\s}^\dag d_{2\s}
+\sum_{\mathbf{k}\s}\en_{{\rm N_2}\mathbf{k}} c^\dag_{{\rm N_2}\mathbf{k}\sigma} c_{{\rm N_2}\mathbf{k}\sigma} 
\nonumber \\ 
&+&\sum_{\mathbf{k}\s} V_{\rm{N_2}\s} \left(d^\dag_{2 \s}c_{\rm{N_2}\mathbf{k}\s} + c_{\rm{N_2}\mathbf{k}\s}^\dag d_{2 \s} \right),
\label{eq:HQD2}
\eea
with standard notation for the electron operators. We assume that the distance between QD$_1$ and QD$_2$ is large enough for their
direct coupling to be negligible. Thus, any cross-correlations induced between the quantum dots will be transmitted by the topological
superconducting nanowire. 

Finally, the low-energy Hamiltonian of the Majorana nanowire can be expressed as \cite{Flensberg-2011}
\be
H_{\rm MW} = \sum_{r=1,2}\sum_{\alpha={\rm L,R}} \sum_\s  (\lambda_{r\s}^{\alpha}d^\dag_{r\s} \eta_\alpha + {\lambda_{r\s}^{\alpha *}} \eta_\alpha 
d_{r\s}) + i \en_M \eta_{\rm L} \eta_{\rm R} ,
\label{eq:HMW}
\ee
where $\eta_\alpha=\eta_\alpha^\dag$ are the self-conjugated Majorana operators,
$\en_M $ stands for an overlap between the Majorana modes, while
$\lambda_{r\s}^{\alpha}$ denotes the hybridization
of $\alpha$-MBS (Majorana bound state) to QD$_{r}$, as illustrated in Fig.\ \ref{scheme}.

It is worth to emphasize, that the Majorana modes hybridize with the side-attached quantum dots, depending on their spin   \cite{Prada-2017,Hoffman-2017,Deng-2018,Yazdani-2018}. This is practically caused by magnetic fields or magnetic textures,
which are necessary to induce the inter-site triplet pairing
in order to allow for topological transition.
To capture such effect, we introduce the following
spin-dependent couplings \cite{Prada-2017}
\be
\label{eq:lambda}
\begin{array}{ll}
\lambda_{1\up}^{L}=\sqrt{2}t_{1}\sin\frac {\theta}{2}, & \hspace{2cm}
\lambda_{1\up}^{R}=-i\sqrt{2}t'_{1} \sin\frac {\theta}{2}, \\
\lambda_{1\down}^{L}=-\sqrt{2}t_{1} \cos\frac {\theta}{2}, &\hspace{2cm}
\lambda_{1\down}^{R}=-i\sqrt{2}t'_{1} \cos\frac {\theta}{2},
\end{array}
\ee
where $t_r$ ($t_r'$) denotes the tunnel matrix elements
between a given quantum dot and neighboring (distant) Majorana mode,
while $\theta$ stands for the spin canting angle characterizing QD$_{1}$-topological nanowire hybrid structure.
Similarly for the other dot,
\be
\label{eq:lambda2}
\begin{array}{ll}
\lambda_{2\up}^{L}=\sqrt{2}t'_{2}\sin\frac {\theta}{2}, & \hspace{2cm}
\lambda_{2\up}^{R}=-i\sqrt{2}t_{2} \sin\frac {\theta}{2}, \\
\lambda_{2\down}^{L}=-\sqrt{2}t'_{2} \cos\frac {\theta}{2}, &\hspace{2cm}
\lambda_{2\down}^{R}=-i\sqrt{2}t_{2} \cos\frac {\theta}{2},
\end{array}
\ee
as illustrated in Fig.\ \ref{scheme}.

Since each quantum dot is also coupled to continuum fermionic states of its own metallic electrode, the in-gap states acquire finite life-times
(for details see Ref.\ \cite{Baranski-2013}).
In the wide bandwidth limit, the resulting broadening is given by the coupling strength
$\G_{{\rm N}_r\s} = \pi \rho_r |V_{{\rm N}_r\s}|^2$,
where $\rho_r$ is the density of states of the given lead,
which is assumed to be constant inside the pairing gap of superconductor.
In addition, the first quantum dot is coupled to the superconducting electrode with coupling strength $\Gamma_S = \pi \rho_S |V_S|^2$, where $\rho_S$ is the density of states of superconductor in the normal state.
For specific calculations, we also impose the spin-independent
couplings, $\G_{{\rm N}_r} = \G_{{\rm N}_r\up}=\G_{{\rm N}_r\down}$,
and use the normal lead band half-width $D$
as convenient unit for the energies, $D \equiv 1$.

The perspectives of using the Majorana modes for quantum computation
rely on their fault-tolerant nature resulting from the topological protection.
To guarantee such protection, the nanowire must be safely longer than the superconducting coherence length, such that the overlap between the Majorana edge modes ($\epsilon_M$) is negligible.
This constraint, however, is hardly satisfied in short nanowires.
Its influence on hybrid structures with a single quantum dot
has been recently considered in Refs.~\cite{Prada-2017,Deng-2018,Ricco-2018, Ricco-2019,Sanches-2020,Gorski-2020}.
Here, we extend these considerations to the setup with 
the second quantum dot defined at the other end of the wire, to allow for nonlocal measurements.

\section{Results}

In experimental realizations of our hybrid structure (Fig.\ \ref{scheme}),
the quantum dots and topological nanowire are usually deposited
directly on (or covered by) superconducting substrate,
while the metallic STM tip is placed in some vicinity of the quantum dots.
It is hence reasonable to expect that the coupling
of ${\rm QD}_1$ to the superconductor is considerably stronger
than the hybridization of individual dots with the external metallic electrodes.
Therefore, for specific computations we assume the following
couplings $\Gamma_S=0.1$ and $\Gamma_{N_1}=\Gamma_{N_2}=0.01$.
To account for a short topological nanowire,
we also impose a finite overlap between the Majorana modes, $\en_M=0.05$. 

In what follows, we present the (normalized) spectral functions of both quantum dots
\begin{eqnarray}
A_{r\sigma}(\omega)= -\Gamma_{N_r} \mbox{\rm Im} \GFzub { d_{r\sigma}}{d_{r\sigma}^{\dag}}_{\omega+i0^{+}} ,
\label{Asig}
\end{eqnarray}
where 
$\GFzub { d_{r\sigma}}{d_{r\sigma}^{\dag}}_{\omega+i0^{+}}$
is the single particle Green’s function obtained exactly from
the equation of motion technique (see the Methods section for details).
Focusing on the nonlocal effects, we neglect the Coulomb repulsion,
however, qualitative effects of such interactions are
briefly discussed in Discussion section.
In the following subsections we consider two qualitatively-different situations: 

\begin{itemize}
\item[{i)}]
The {\it local setup}, where the topological segment is long enough to prohibit a direct tunneling of electrons from the given quantum dot to the opposite-side Majorana mode. In other words, in this scenario each quantum dot is assumed to be coupled only to its neighbouring edge state, i.e. $t'_{1}=t'_{2}=0$ and thus $\lambda_{1\sigma}^R = \lambda_{2\sigma}^L = 0$.  
\item[{ii)}]
 The {\it nonlocal setup}, in which the electrons can be exchanged between the quantum dots and both boundary Majorana modes, but their couplings are pretty asymmetric.
 It is reasonable to assume that even for a very short wire, the tunnel coupling between each quantum dot and its neighbouring edge state is much stronger than the coupling to the edge state on the opposite side of the chain. 
 We thus assume $t_{1}=t_{2}=0.05$, unless stated otherwise, and study how the system's characteristics depend on the coupling of each quantum dot
 to the corresponding distant Majorana mode, $t'_{1}=t'_{2}\neq 0$.
\end{itemize}
%

\subsection{The case of local correlations}
\label{sec:resultslin}
\subsubsection*{Quasiparticle states and spectral behavior}

\begin{figure*}
		\centering
		\includegraphics[width=0.8\linewidth]{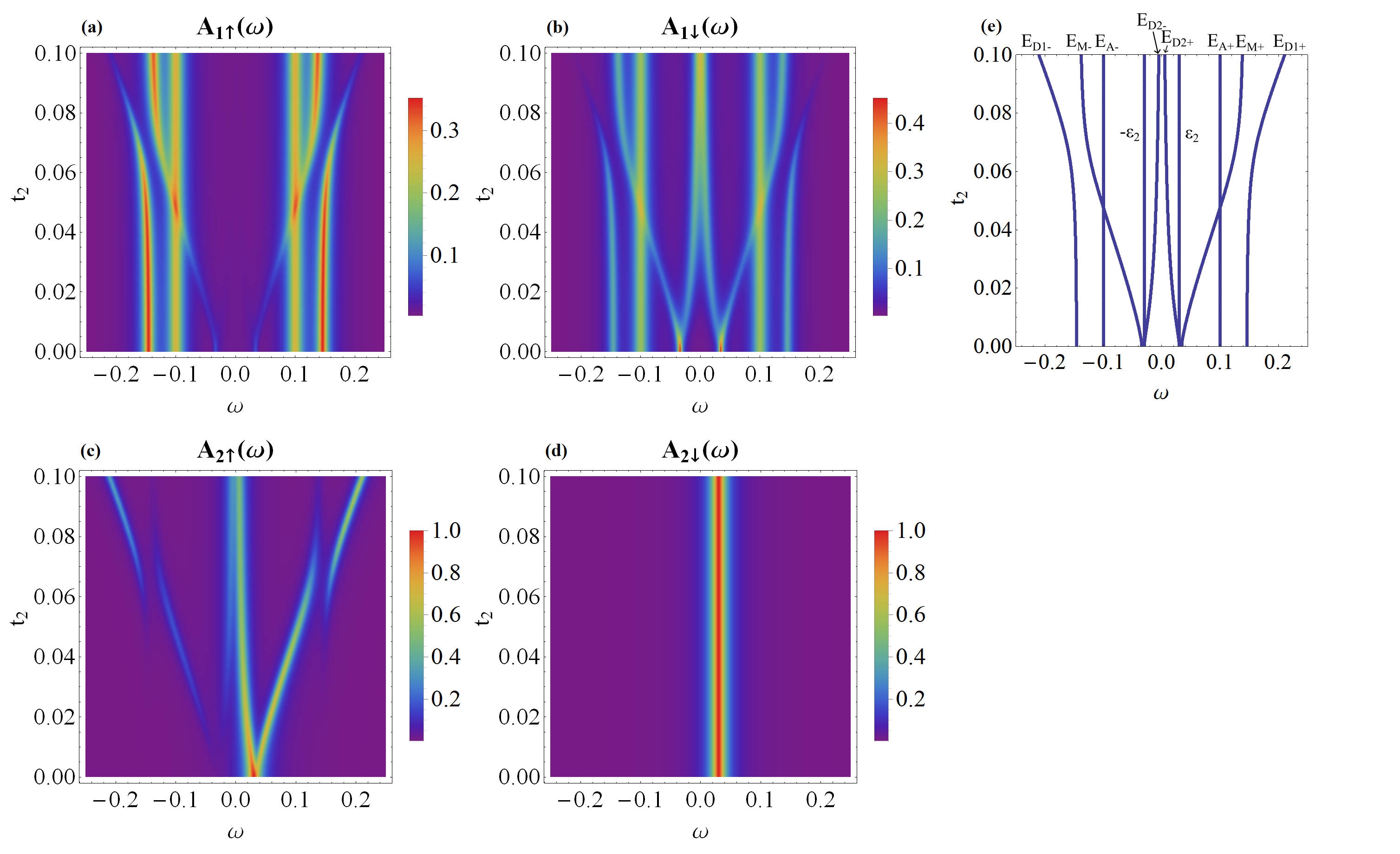}
		\caption{The energy dependence of the spin-resolved spectral function of QD$_{1}$ (a and b panels) and QD$_{2}$ (c and d panels) versus the coupling $t_{2}$.
        Results are obtained for the following model parameters:
        $\epsilon_1=0$, $\epsilon_2=0.03$, $\en_M=0.05$, $t_{1}=0.05$ and $\theta=\pi$. Panel (e) displays the quasiparticle energies $E_{M\pm}$, $E_{D1\pm}$ and $E_{D2\pm}$ obtained for $\Gamma_{N1}=0=\Gamma_{N2}$.}
	\label{Figtm01q2}
\end{figure*}

To begin with, let us analyze the local case, i.e.
when quantum dots are directly coupled only to the neighboring Majorana modes. 
It is instructive to investigate the specific situation when only QD$_1$ is coupled to MW ($t_{1}\neq 0$), 
ignoring any coupling of QD$_2$ to the topological superconductor ($t_{2} = 0$).
Under such conditions, an interplay of the on-dot pairing and
the leakage of the Majorana mode gives rise to emergence of $3$ pairs of quasiparticles. 
Two of them occur at $E_{A\pm}=\pm E_{A}$,
where $E_A=\sqrt{\epsilon_1^2+\Gamma_S^2}$, and represent the conventional 
Andreev bound states \cite{Baranski-2017, Gorski-2018, Zienkiewicz_2020}.
The additional quasiparticles at energies
\bea
E^0_{M1\pm}&=&\pm\sqrt{A-sgn(E_A-\epsilon_M)\sqrt{A^2-\epsilon_M^2 E_A^2}},
\label{EM_state}\\
E^0_{D1\pm}&=&\pm\sqrt{A+sgn(E_A-\epsilon_M)\sqrt{A^2-\epsilon_M^2 E_A^2}} ,
\label{ED_state}
\eea
where $A=2t_{1}^2+\frac{\epsilon_M^2+E_A^2}{2}$, originate from the Majorana modes and depend on the overlap $\epsilon_M$ and $sgn(x)$ is signum function.
For a long topological chain ($\epsilon_M \rightarrow 0$), two of these quasiparticle states (\ref{EM_state}) merge into a zero-energy level, $E_{M1 \pm}=0$.
Then, the spectrum of QD$_1$ consists of: $2$ usual Andreev states,
$2$ Andreev-Majorana hybrid quasiparticles and the (doubly degenerate) zero-energy mode. 
Due to the superconducting proximity effect, all these quasiparticle states 
contribute to the spectral density of QD$_1$,
though with different spectral weights depending on spin,
cf.~\figs{Figtm01q2}(a) and \ref{Figtm01q2}(b) for $t_2=0$.
Some more details about this Majorana leakage in the limit of $\epsilon_M\rightarrow 0$ have been discussed in Ref.\ \cite{Zienkiewicz_2020}.

In the opposite case, when only QD$_2$ is coupled to the topological nanowire
($t_{2} \neq 0$) and QD$_1$ is absent ($t_{1} = 0$), 
one obtains $5$ quasiparticle states. 
Their character, however, is very different from the former situation,
due to the absence of the Andreev bound states.
The ground state of QD$_{2}$ occurs near $\epsilon_2$,
which is neither affected by $t_{2}$ nor by $\epsilon_M$. 
Its hole-counterpart at $-\epsilon_2$ has no spectral weight in
the absence of conventional superconducting lead.
The remaining Majorana-like $E^0_{M2\pm}$ and dot-like $E^0_{D2\pm}$
quasiparticle levels emerge at
\bea
E^0_{M2\pm}=\pm\sqrt{B-sgn(\epsilon_2-\epsilon_M)\sqrt{B^2-\epsilon_M^2 \epsilon_2^2}},
\label{EM2pm}\\
E^0_{D2\pm}=\pm\sqrt{B+sgn(\epsilon_2-\epsilon_M)\sqrt{B^2-\epsilon_M^2 \epsilon_2^2}} ,
\label{ED2pm}
\eea
where $B=2t_{2}^2+\frac{\epsilon_M^2+\epsilon_2^2}{2}$,
in analogy to Eqs.~(\ref{EM_state}) and (\ref{ED_state}). 

For $t_{1} \neq 0$ and $t_{2} \neq 0$, 
the canting intertwines
the spectra of QD$_1$ and QD$_2$, 
even when the Majorana mode couples
only to the spin-$\uparrow$ channel ($\theta=\pi$),
as presented in \fig{Figtm01q2} for $t_2 > 0$. 
The quasiparticles at QD$_1$ include two Andreev-type bound states $E_{A\pm}$ 
independently of the coupling of QD$_2$ to MW.
Similarly, for QD$_2$, we observe the state at $\epsilon_{2}$,
no matter what the coupling of QD$_1$ to topological superconductor is. 
The remaining quasiparticles arise from the hybridization
of QD$_1$ and QD$_2$ states with the Majorana mode ($E_{M\pm}$, $E_{D1\pm}$ and $E_{D2\pm}$).

\begin{figure}
		\centering
		\includegraphics[width=0.48\textwidth]{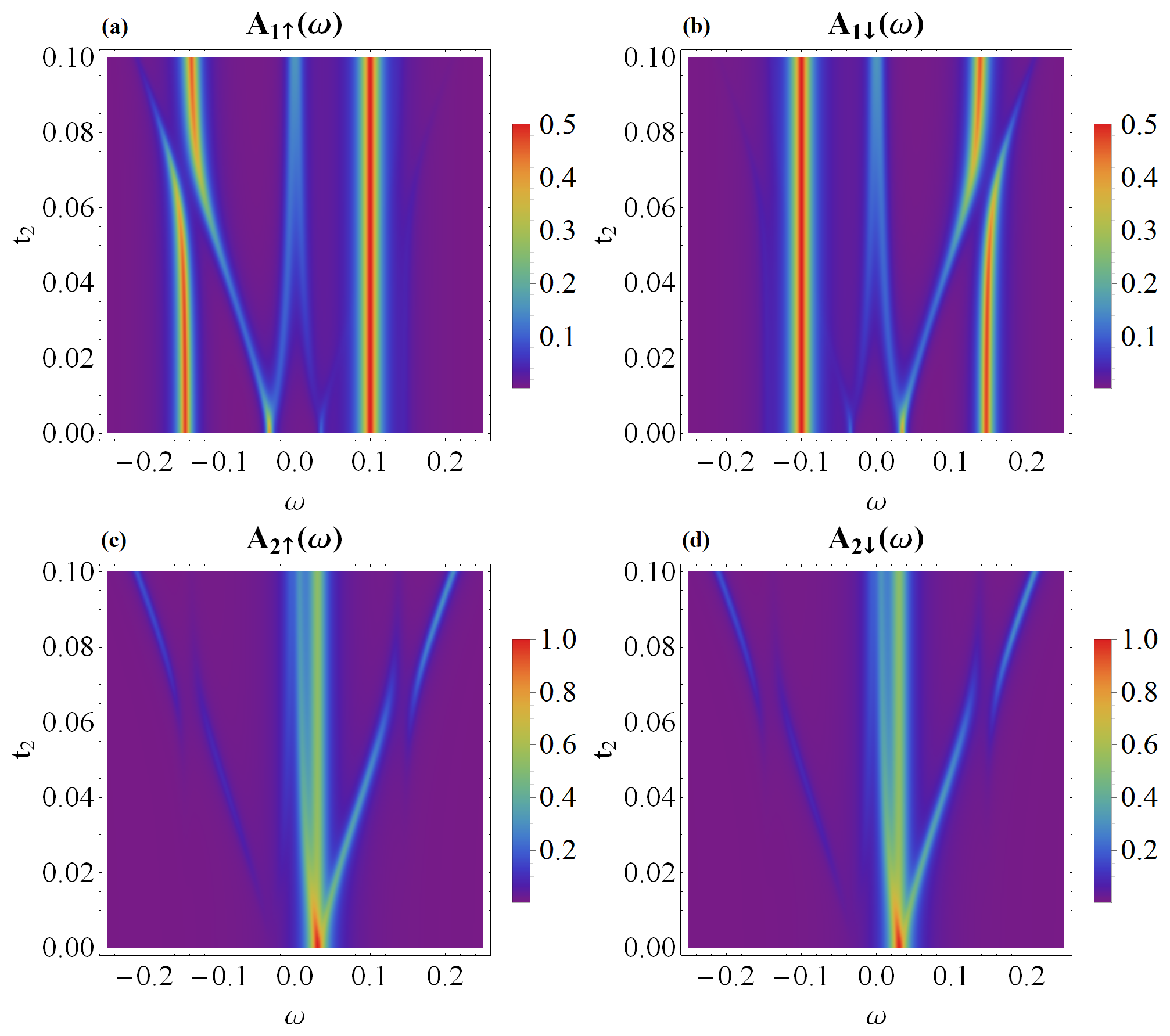}
			\caption{Variation of the spin-dependent spectral functions of QD$_{1}$ (a and b panels) and QD$_2$ (c and d panels) with respect to the coupling $t_{2}$
   calculated for the same model parameters as in Fig.~\ref{Figtm01q2}
   assuming the spin canting angle $\theta=\frac{\pi}{2}$. The quasiparticle energies $E_{M\pm}$, $E_{D1\pm}$ and $E_{D2\pm}$ coincide with those presented in Fig. \ref{Figtm01q2} (e). 
   }
	\label{Figtm01q2th05Pi}
\end{figure}

Figure \ref{Figtm01q2}, by presenting the spin-resolved spectral
functions of individual quantum dots, $A_{r\sigma}(\omega)$,
reveals the influence of the coupling $t_{2}$ between QD$_2$ and the Majorana mode on the energy spectrum of QD$_1$ and QD$_2$, respectively. 
In $A_{1 \uparrow}(\omega)$ its effect relies mainly on shifting the energy levels of the hybrid quasiparticles. 
More interesting results can be seen due to the relationship between the energy spectrum of spin-$\downarrow$ electrons of QD$_1$ and spin-$\uparrow$ electrons of QD$_2$. 
If $t_2$ is non-vanishing, one of the hybrid states in $A_{1 \downarrow}(\omega)$ splits to form the high and low energy branches. 
In total, we obtain $8$ quasiparticle states associated with:
$2$ ABSs and $6$ hybrid states. 
The molecular states appearing in $A_{2 \uparrow}(\omega)$ coincide with the energies of molecular states in $A_{1 \downarrow}(\omega)$. 
Such result indicates a mutual origin of the opposite spin electrons
at the distant quantum dots transmitted through the Majorana modes. 
Moreover, Fig.\ \ref{Figtm01q2}(b) clearly shows that $A_{1\downarrow}(\omega)$
is indirectly affected by the topological 
superconductor even when only the spin-$\uparrow$ electrons are directly coupled to the wire.
This is caused by the on-dot pairing induced by the superconducting lead \cite{Gorski-2018}. 
On the contrary, the spectral function $A_{2\downarrow}(\omega)$
is then unaffected by the presence of the Majorana modes, cf.~\fig{Figtm01q2}(d). 

We now consider the case when the canting leads to noncolinearity
by setting $\theta=\frac{\pi}{2}$, cf.~Eqs.~(\ref{eq:lambda}) and (\ref{eq:lambda2}). 
The corresponding spin-dependent spectral functions
for this situation are presented in Fig.\ \ref{Figtm01q2th05Pi}.
First of all, one can clearly notice that the spectrum of QD$_1$ is characterized by $A_{1\up}(\w) = A_{1\down}(-\w)$. 
Furthermore, $A_{1\sigma}(\w)$ reveals only a single Andreev peak, at $\omega = E_{A+}$ for 
spin-$\uparrow$ and at $\omega=E_{A-}$ 
for spin-$\downarrow$ sectors, respectively.
Additionally, in both spectral functions $A_{1\sigma}(\w)$ we recognize 
the quasiparticle peaks at $E_{M\pm}$ originating from QD$_2$.
The symmetry of the spin-resolved spectral functions $A_{1\sigma}(\w)$ displayed in Fig.~\ref{Figtm01q2th05Pi}  originates partly from the canting angle $\theta=\pi/2$ and partly from the superconducting proximity effect, which mixes the particle with hole degrees of freedom. Physically it means that
the spin-$\uparrow$ electrons ($\omega<0$) are mixed with spin-$\downarrow$ holes ($\omega>0$).
Similar effect occurs between $\omega<0$ spin-$\downarrow$ and $\omega>0$
spin-$\uparrow$ quasiparticle states of QD$_{1}$.
Concerning the spectral behavior of QD$_2$, we observe that $A_{2\up}(\w)=A_{2\down}(\w)$,
and in both spin sectors there are present the quasiparticles at $\omega=E_{D1\pm}$,
transmitted from QD$_1$ via the topological superconducting nanowire. 

\subsubsection{Andreev transmission}

Junctions consisting of metallic and superconducting leads allow for the anomalous electron-to-hole (Andreev) scattering mechanism, 
active particularly in the subgap regime. 
In the N$_1$-QD$_1$-SC circuit we can observe the direct Andreev reflection (DAR) processes, when the incoming electron from the metallic lead is scattered as a hole back to the same electrode. In addition, the side-attached topological superconductor allows for the non-local crossed Andreev reflection (CAR), in which a hole is nonlocally transmitted to the electrode 
on the right hand side of the wire \cite{Nilsson-2008,Law-2009,Liu-2013,He-2014}.
The transmission coefficients for the direct and crossed Andreev refection processes 
can be expressed by the off-diagonal terms of the matrix Green's functions
%
%
%
\bea
T_r^{DAR}(\omega)=\sum_{\s}\Gamma_{N_r\s}\Gamma_{N_r\bar{\s}}\left|
\GFzub{d_{r\s}}{d_{r\bar{\s}}}_\omega
\right|^{2} ,
\label{eq:TDAR}
\eea
and
\bea
T^{CAR}(\omega)&=&\sum_{\s}\big(\Gamma_{N_1\s}\Gamma_{N_2\bar{\s}}\left|
\GFzub{d_{1\s}}{d_{2\bar{\s}}}_\omega\right|^{2}
\nonumber \\
&+&\Gamma_{N_2\s}\Gamma_{N_1\bar{\s}}\left|
\GFzub{d_{2\s}}{d_{1\bar{\s}}}_\omega
\right|^{2}\big) .
\label{eq:TCAR}
\eea

\begin{figure*}
		\centering
		\includegraphics[width=0.85\textwidth]{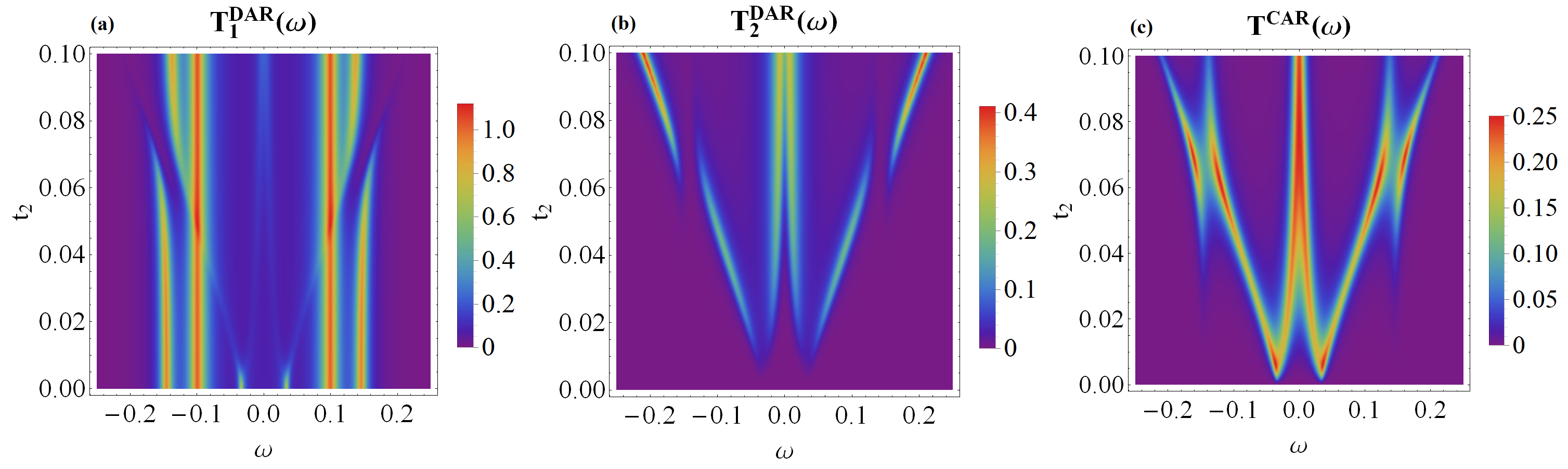}
	\caption{The transmittance of the DAR and CAR tunneling channels plotted 
        versus energy $\omega$ and the coupling $t_{2}$ assuming the spin canting angle
        $\theta=\frac{\pi}{2}$ and the other parameters the same as in Fig.~\ref{Figtm01q2}.}
	\label{FigTDARTCARlin}
\end{figure*}

Figure~\ref{FigTDARTCARlin} displays the transmittance obtained for the direct (local) and crossed (nonlocal) Andreev tunneling. In the low-temperature limit, the differential tunnelling conductance would be a product of the transmittance and conductance quantum $4e^{2}/h$. By inspecting the DAR channel, we find signatures
of all quasiparticle states induced in our hybrid structure. For $\theta=\pi$, the second quantum dot neither reveals the direct nor crossed particle-to-hole reflections
because superconducting proximity effect is absent. For $\theta=\pi/2$ (as well as
for other angles different from multiples of $\pi$) we clearly see the nonlocal
(CAR) reflections and even the direct Andreev scattering in the N$_{2}$-QD$_{2}$-MW setup.

%
%
%
%

\subsection{Nonlocal correlations}
\label{sec:resultspel}
\subsubsection*{Spectral behavior and Andreev transmittance}

Let us now focus on the case when the quantum dots side-attached to the short topological superconductor hybridize with 
both boundary Majorana modes,
$t'_{r}\neq 0$, and the ratio $\eta^2=\left|t'_{r}/t_{r}\right|$ 
can be regarded as a qualitative measure  of  nonlocality in this configuration 
\cite{Prada-2017,Clarke-2017,Deng-2018,Ricco-2019}. 
Figures\ \ref{Fig_A12_tprim}(a) and (b) show the electronic spectrum of QD$_{1}$, 
while varying with respect to the nonlocal coupling $t'_{1}=t'_{2}$. We notice substantial influence 
on the Andreev states $E_A$  accompanied by redistribution of their spectral weights. 
In particular, the quasiparticle state at $E_{A-}$ of the spin-$\uparrow$ sector transmits 
its spectral weight to the state at $E_{D1-}$. Similarly, in the spin-$\downarrow$ sector 
the spectral weight of $E_{A+}$ is transferred to $E_{D1+}$. Furthermore, we observe 
additional changes in the spectrum of QD$_1$ appearing
at energies $E_{D2\pm}$, originating from 
QD$_2$ transmitted via the short topological nanowire.

%

\begin{figure}[tbh!]
		\centering
		\includegraphics[width=0.48\textwidth]{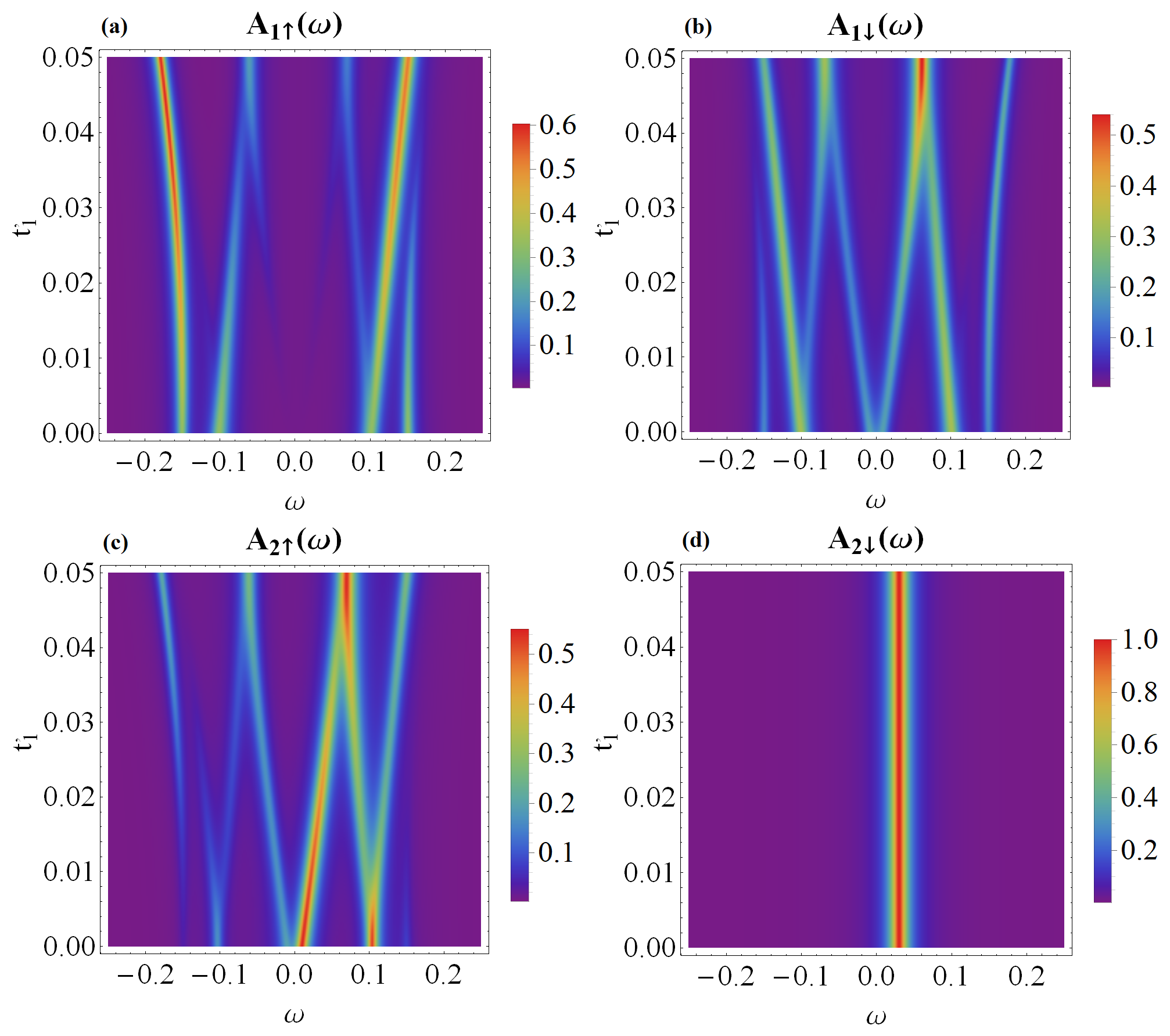}
	\caption{Evolution of the spin-resolved spectra of QD$_1$ (panels a, b) and QD$_2$ (panels c, d) obtained for finite nonlocal couplings $t'_{1}=t'_{2}$. 
 The parameters are the same as in 
 Fig.~\ref{Figtm01q2} with $t_{1}=t_{2}=0.05$.}
	\label{Fig_A12_tprim}
\end{figure}

\begin{figure}[b!]
		\centering
		\includegraphics[width=0.48\textwidth]{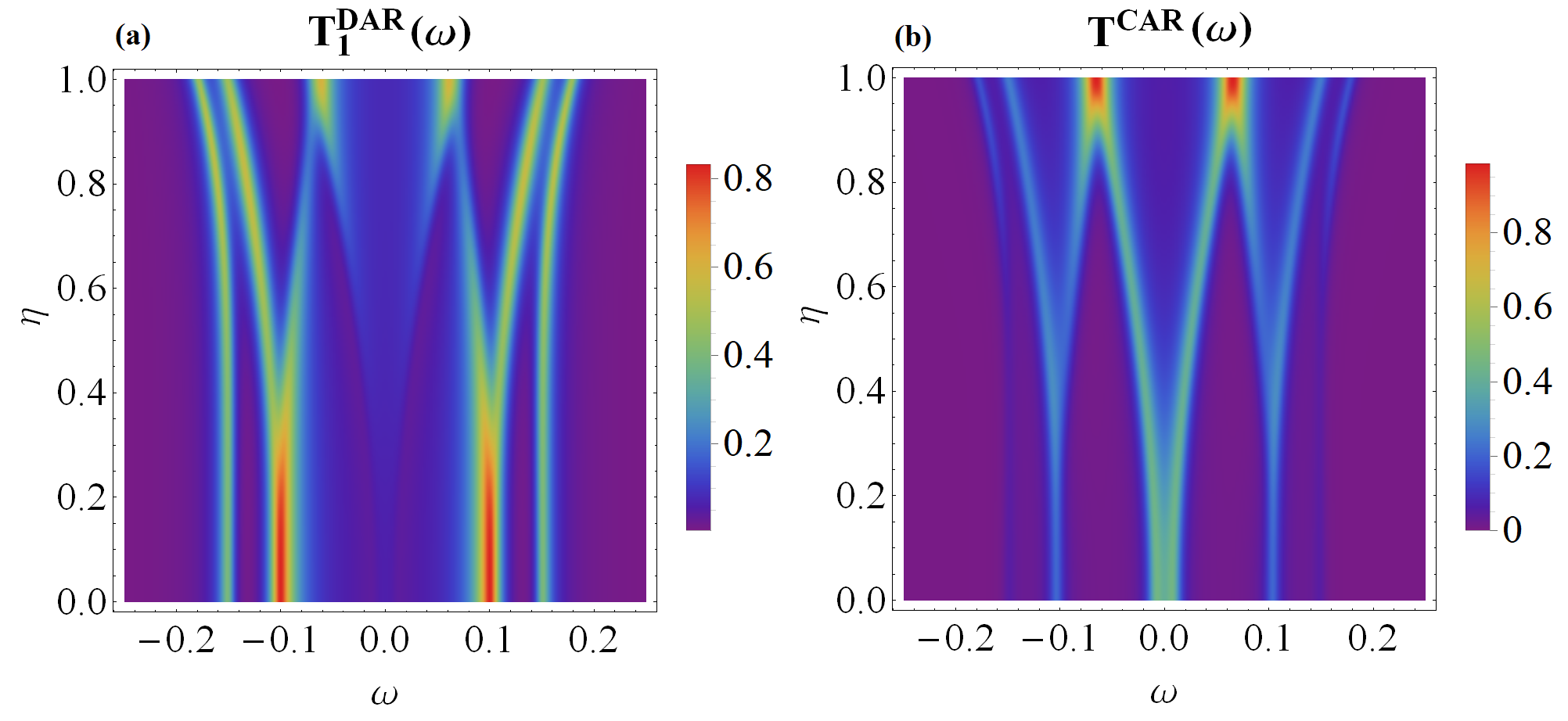}
	\caption{Transmittance of the DAR and CAR transport channels versus energy $\omega$ and the MBS nonlocality parameter $\eta=\sqrt{\left|t'_{r}/t_{r}\right|}$. Results are obtained for the same parameters as in Fig.~\ref{Fig_A12_tprim}.
 }
	\label{Fig_TA_tprim}
\end{figure}

The spin-resolved spectral functions of QD$_2$, revealing the corresponding quasiparticle states, are shown in Figs.\ \ref{Fig_A12_tprim}(c) and (d).
For the colinear case, $\theta=\pi$, we clearly see that only spin-$\uparrow$ sector 
of QD$_{2}$ exhibits the same quasiparticle states as  QD$_{1}$. The opposite spin is 
completely unaffected, as manifested by a single peak at $\varepsilon_{2}$.
%
Such a behavior indicates absence of any on-dot pairing at QD$_2$. For other values of $\theta$,
the spin-resolved spectral function of QD$_2$ shows signatures of the conventional quasiparticle 
states ($E_{D1\pm}$) and Majorana-type features ($E_{M\pm}$) acquired from QD$_1$ via the short 
topological superconductor. For $t'_{1} \neq 0$, the Andreev bound states become mixed with the 
quasiparticle states $E_{D1\pm}$, therefore
weak superconducting correlations are indirectly 
induced at QD$_2$.  

Figure \ref{Fig_TA_tprim} presents the signatures of the quasiparticle states possible to be probed
by the Andreev spectroscopy. Specifically, we show variation of the direct and crossed Andreev 
transmittance with respect to $\eta=\sqrt{\left|t'_{r}/t_{r}\right|}$ obtained for  the symmetric couplings 
$t_1=t_2$ and $t'_{1}=t'_{2}$. We notice that the local Andreev scattering (DAR) is a particle-hole 
symmetrized version of the spin-resolved spectral functions of QD$_{1}$ [see panel a and b in Fig.\ 
\ref{Fig_A12_tprim}]. The nonlocal (CAR) transmittance, on the other hand, mixes the particle with 
hole degrees of freedom between the distant quantum dots. Such effect is partly caused by the overlap 
of Majorana modes, $\epsilon_{M}\neq 0$, and additionally comes from the nonlocal hopping, $t'_{r}\neq 0$.

\subsubsection{Singlet and triplet electron pairing}

\begin{figure*}
\centerline{\includegraphics[width=0.85\linewidth]{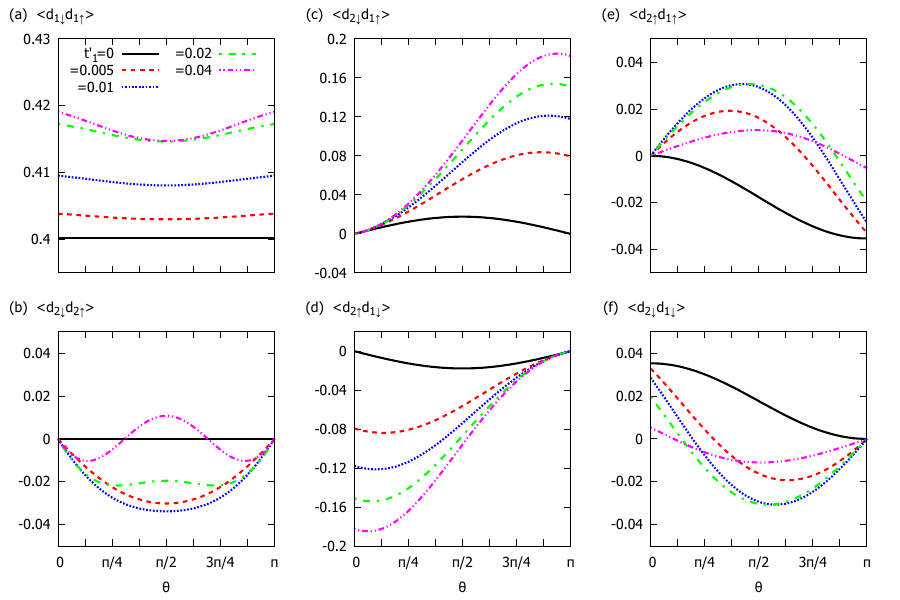}}
\caption{Various channels of the electron pairing induced at individual quantum dots and nonlocally between the dots versus the spin canting angle, $\theta$. Results are obtained at zero temperature for several values of the coupling $t'_{1}=t'_{2}$ (as indicated), assuming the following model parameters: 
$\epsilon_1=\epsilon_2=0$, $\en_M=0.05$ and $t_{1}=t_{2}=0.05$.}
\label{Fig_n1458_theta}
\end{figure*}

To characterize the efficiency of the superconducting proximity effect we have calculated the expectation values 
of the on-dot electron pairings, which (at zero temperature) are defined by
\bea
\langle d_{r\down}d_{r\up}\rangle=-\frac{1}{\pi}\int_{-\infty}^0{\mbox{\rm Im} 
\GFzub{d_{r\up}}{d_{r\down}}_\omega \; d\omega}.
\label{nd14}
\eea
In addition, we have also studied the nonlocal electron pairs induced between the quantum dots, considering the singlet and triplet channels   \cite{Su-2017}.
%
%
For the local situation, $t_{i}'=0$, the local pairing induced at QD$_{1}$ is insensitive to the spin canting angle, whereas the on-dot pairing of QD$_{2}$ is absent for all values of $\theta$. In contrast to this, for $t'_{1}\neq 0$, the Andreev bound states $E_{D1\pm}$ appear simultaneously in both quantum dots, though with different amplitudes. The local pairings depend then on both $t'_{1}$ and the spin canting angle $\theta$.
In Fig.~\ref{Fig_n1458_theta}(a) we present the dependence of the local pairing $\langle d_{1\down}d_{1\up}\rangle$ on $\theta$ for several values of the hopping $t'_{1}=t'_{2}$, as indicated. Optimal conditions for this pairing coincide with the colinear cases ($\theta=0$ or $\theta=\pi$) and the minimal local pairing occurs at perpendicular canting, $\theta=\pi/2$. 
Figure\ \ref{Fig_n1458_theta}(b) illustrates the variation of the on-dot pairing of QD$_2$ with respect to $\theta$. In the weak coupling limit (small $t'_{1}$), the local electron pairing induced in QD$_2$ is negative and its minimum occurs at $\theta=\pi/2$. 
On the other hand, in the strong coupling limit (for large $t'_{1}$), such indirectly induced electron pairing changes sign and its maximum occurs again at $\theta=\pi/2$. 
For the colinear cases ($\theta=0$ or $\theta=\pi$), the local pairing of QD$_{2}$ vanishes, $\langle d_{2\down}d_{2\up}\rangle=0$, no matter what the value of $t'_{1}$ is.

To summarize this section, we emphasize that the short topological superconductor with a finite overlap between the Majorana modes ($\epsilon_M \neq 0$) enables crossed Andreev reflections, originating from the nonlocal electron pairing $\langle d_{2\sigma}d_{1\sigma'}\rangle$.
For the colinear cases, we observe vanishing of the nonlocal  correlations $\langle d_{2 \uparrow} d_{1 \uparrow} \rangle$ and $\langle d_{2 \downarrow} d_{1 \uparrow} \rangle$ (for $\theta=0$, i.e. no spin canting) or pairings $\langle d_{2 \uparrow} d_{1 \downarrow} \rangle$ and $\langle d_{2 \downarrow} d_{1 \downarrow} \rangle$ (for $\theta=\pi$, i.e. spin quantization axis reversal between the wire ends).
Otherwise, all the inter-dot electron pairs survive, allowing for the nonlocal Andreev scattering.
Detailed dependence of the inter-dot (singlet and triplet) pairing functions against the spin canting angle ($\theta$) is displayed in Figs.\ \ref{Fig_n1458_theta}(c)-(f). 



\section{Discussion}
\label{sec:Discussion}

We have investigated the quasiparticle spectrum 
of the hybrid structure formed of two quantum dots interconnected 
through a short topological superconducting nanowire, hosting 
the overlaping Majorana boundary modes.
We have found clear signatures of nonlocal (inter-dot) correlations
in spectral functions of the system, which would be feasible for detection with STM techniques.
Their precise form depends on the magnitude of the dots coupling 
to the boundary modes of the topological wire.
For the case when quantum dots are coupled only to the adjacent edges 
of the wire, we predict the following manifestations of nonlocality.
\begin{itemize}
\item[(i)]{In maximally spin-canted case ($\theta=\pi$), 
the spin-$\downarrow$ QD$_1$ spectrum gains contributions from all the
quasiparticles visible in the spin-$\uparrow$ QD$_2$ spectrum,  cf.~Fig.~\ref{Figtm01q2}.
This can be seen as transmission of the quasiparticle weight through
 the Majorana modes.
At the same time, only some of the QD$_1$ quasiparticles 
are admitted to QD$_2$, which is best illustrated by 
the spin-$\downarrow$ QD$_2$ spectrum, unaffected by coupling to the wire.}


\item[(ii)]{For generic canting, $\theta\neq 0$ and $\theta\neq \pi$, the complete set of the molecular quasiparticles  are apparent in the spin-resolved speactral functions of both quantum dots.}
\end{itemize}
In more general case, when the dots are coupled to both Majorana modes ($t'_1=t'_2\neq 0$), the  inter-dot electron pairs are formed, both in the singlet and triplet channels.  Such nonlocal pairs can be experimentally detected by the crossed Andreev reflection spectroscopy, using either the unpolarized or spin-polarized external electrodes.

In the present work we consider the case of quantum dots 
energy levels close to the Fermi energy.
For further in-depth study it would be useful to take into account the 
on-site Coulomb repulsion between the opposite-spin electrons.
This would open the way to address also Coulomb blockade and Kondo regimes.
Previous studies of the superconducting nanohybrid structures indicated 
that the strong Coulomb repulsion could also suppress the local pairing potential. 
For this reason one might expect the correlations to predominantly affect 
the quaspiarticle energies of the bound states at QD$_1$. As regards the second
quantum dot, the correlations would give rise to the Zeeman term \cite{Lutchyn-2014,Rosa_Lopez-2013}.
Additionally, in both quantum dots renormalization of the effective spin exchange 
interaction would occur, playing essential role in the low-temperature Kondo regime, 
when the quantum dots are approximately half-filled.
These issues, however, are beyond the scope of the present 
study, which is focused on the nonlocal pairing correlations 
relevant for the crossed Andreev spectroscopy.

Let us finally comment on possible experimental means of verifying our predictions.
T. Dvir et al. \cite{Kouewenhoven-2023} have provided experimental evidence for the realization of
the Kitaev chain composed of just two sites. In practice, these sites represented
pieces of the semiconducting nanowire with discrete energy levels brought in contact via the conventional superconductor. Depending on the energy levels
(tunable by gate potentials) and affected by the spin-orbit and  Zeeman effect
these sites were able to approach sweet points, inducing the Majorana
modes. Since they were spatially close to one another, their wave functions might have been hybridized ($\epsilon_{M}\neq 0$). Our present study
could be relevant to this situation with additional quantum dots 
attached on opposite sides. Upon forming the circuits with external electrodes,
the quasiparticle states of these attached quantum dots can be probed experimentally.
For low bias voltages (smaller than the energy gap of superconductor), the only transport
channel would by contributed by the particle-to-hole (Andreev) scattering. Our
estimations of the Andreev transmittance (corresponding roughly to the differential
Andreev conductance at low temperatures) would thus enable exploring the direct and crossed Andreev conductance, revealing the local and nonlocal cross-correlations transmitted between these dots via the overlapping Majorana modes.


\section{Methods}
\label{sec:Methods}

To study the hybrid structure shown in Fig.\ \ref{scheme},
we have determined the quasiparticle states. We have taken into 
account the leakage of Majorana modes into both quantum dots and the superconducting 
proximity effect, the latter inducing conventional Andreev 
bound states at QD$_{1}$. 
Focusing on the subgap energy region, $|\omega| \ll \Delta_{S}$, we simplified 
the considerations by treating $\Delta_{S}$ as the largest energy scale. In the 
limit of $\Delta_S \rightarrow \infty$, the fermionic degrees of freedom of the 
superconducting lead can be integrated out, giving rise to the on-dot electron 
pairing $- \Gamma_{S} (d_{1 \up}^\dag d_{1 \down}^\dag + d_{1 \down} d_{1 \up})$. 
Under such circumstances the “proximitized” QD$_1$ can be modeled by \cite{Bauer-2007,Yamada-2011,Baranski-2013,Gorski-2018}
\begin{widetext}
\bea \label{eq:HQD1ef}
H_{\rm L} = \sum_{\s} \en_{1\sigma} d_{1\s}^\dag d_{1\s}
 - \Gamma_{S} (d_{1 \up}^\dag d_{1 \down}^\dag + d_{1 \down} d_{1 \up})+\sum_{\mathbf{k}\s}\en_{{\rm N_1}\mathbf{k}} c^\dag_{{\rm N_1}\mathbf{k}\sigma} c_{{\rm N_1}\mathbf{k}\sigma}
+\sum_{\mathbf{k}\s} V_{\rm{N_1}\s} \left(d^\dag_{1 \s}
c_{\rm{N_1}\mathbf{k}\s} + c^\dag_{\rm{N_1}\mathbf{k}\s} d_{1 \s} \right).
\eea
It is also convenient to recast the Majorana operators by the usual fermion operators, $\eta_L = (f^\dag+f)/\sqrt{2}$ and $\eta_R = i(f^\dag-f)/\sqrt{2}$. In this representation, the term (\ref{eq:HMW}) of the model Hamiltonian can be expressed as
\bea \label{eq:HMWef}
H_{\rm MW} = \sum_{r=1,2}\sum_\s \left[ t_{r\s}^+\left(d_{r\s}^\dag f^\dag+f d_{r\s}\right)+t_{r\s}^-\left(d_{r\s}^\dag f+f^\dag d_{r\s}\right)\right]+\en_M\left(f^\dag f-\frac{1}{2}\right) ,
\eea
with the spin-dependent hopping integrals
\be
\label{eq:tpm1}
\begin{array}{cc}
t_{1\up}^+=t_{1}\sin\frac {\theta_L}{2}+t'_{1}\sin\frac {\theta_R}{2} , 
&t_{1\up}^-=t_{1}\sin\frac {\theta_L}{2}-t'_{1}\sin\frac {\theta_R}{2} , \\
t_{1\down}^+=-t_{1}\cos\frac {\theta_L}{2}+t'_{1}\cos\frac {\theta_R}{2} , 
&t_{1\down}^-=-t_{1}\cos\frac {\theta_L}{2}-t'_{1}\cos\frac {\theta_R}{2} , 
\end{array}
\ee
and
\be
\label{eq:tpm2}
\begin{array}{cc}
t_{2\up}^+=t'_{2}\sin\frac {\theta_L}{2}+t_{2}\sin\frac {\theta_R}{2} , 
&t_{2\up}^-=t'_{2}\sin\frac {\theta_L}{2}-t_{2}\sin\frac {\theta_R}{2}, \\
t_{2\down}^+=-t'_{2}\cos\frac {\theta_L}{2}+t_{2}\cos\frac {\theta_R}{2}, 
& t_{2\down}^-=-t'_{2}\cos\frac {\theta_L}{2}-t_{2}\cos\frac {\theta_R}{2}.
\end{array}
\ee
The influence of the topological nanowire on the side attached quantum dots can be
analyzed within the Green's function approach represented in the particle-hole 
matrix notation, 
$\hat {\cal{G}}(\omega )=
\GFzub{\Psi}{\Psi^\dag}_{\omega}$,
with the Nambu spinor defined as $\Psi  = (d_{1\uparrow},d_{1\uparrow}^\dag,d_{1\downarrow},d_{1\downarrow}^\dag,d_{2\uparrow},d_{2\uparrow}^\dag,d_{2\downarrow},d_{2\downarrow}^\dag,f,f^\dag)$.
The equation of motion
\begin{eqnarray}
(\omega+i 0^{+}) \GFzub{\Psi_i}{\Psi_j}_\omega= \left\langle \left[\Psi_i,\Psi_j\right]_+ \right\rangle + \GFzub{\left[\Psi_i,H\right]_-}{\Psi_j}_\omega
\label{EOM}
\end{eqnarray}
yields the following retarded Green's function 
%
\begin{eqnarray}
{\cal{G}}^{-1}(\omega) =\omega \hat{I}+
\left( \begin{array}{cccccccccc}
-\tilde{\epsilon}_{1\uparrow} &0&0& \Gamma_{S}& 0 & 0& 0 & 0 & -t^{-}_{1\up} & -t^{+}_{1\up}\\
0&(\tilde{\epsilon}_{1\uparrow})^{*}&-\Gamma_{S}& 0 & 0 & 0& 0 & 0& t^{+}_{1\up}& t^{-}_{1\up}\\
0&-\Gamma_{S}&-\tilde{\epsilon}_{1\downarrow}& 0 & 0 & 0& 0 & 0&-t^{-}_{1\down} & -t^{+}_{1\down}\\
\Gamma_{S}&0&0&(\tilde{\epsilon}_{1\downarrow})^* & 0 & 0& 0 & 0& t^{+}_{1\down} & t^{-}_{1\down}\\
0 & 0& 0 & &-\tilde{\epsilon}_{2\uparrow}&0&0& 0&-t^{-}_{2\up} & -t^{+}_{2\up} \\
0 & 0& 0 & 0&0&(\tilde{\epsilon}_{2\uparrow})^*&0& 0& t^{+}_{2\up} & t^{-}_{2\up}\\
0 & 0& 0 & 0&0&0&-\tilde{\epsilon}_{2\downarrow}& 0&  -t^{-}_{2\down} & -t^{+}_{2\down}\\
0& 0& 0 & 0&0&0&0&(\tilde{\epsilon}_{2\downarrow})^*&t^{+}_{2\down} & t^{-}_{2\down} \\
-t^{-}_{1\up} & t^{+}_{1\up} &-t^{-}_{1\down} & t^{+}_{1\down}& -t^{-}_{2\up} & t^{+}_{2\up} &-t^{-}_{2\down} & t^{+}_{2\down}& -\epsilon_{M}& 0\\
-t^{+}_{1\up}& t^{-}_{1\up} &-t^{+}_{1\down} & t^{-}_{1\down}& -t^{+}_{2\up} & t^{-}_{2\up} &-t^{+}_{2\down} & t^{-}_{2\down}& 0    & \epsilon_{M}
\end{array}\right) ,
\label{Green1010}
\end{eqnarray}
%
where $\hat{I}$ stands for the identity matrix and $\tilde{\epsilon}_{i\sigma}=\epsilon_{i\sigma}-i\Gamma_{N_i}$.
We have used various parts of this matrix Green's function for computing the spin-resolved spectral functions of both quantum dots,
the direct and crossed Andreev transmittances as well as for
evaluation of the local and nonlocal pairing functions.

\begin{figure}
		\centering
		\includegraphics[width=0.8\textwidth]{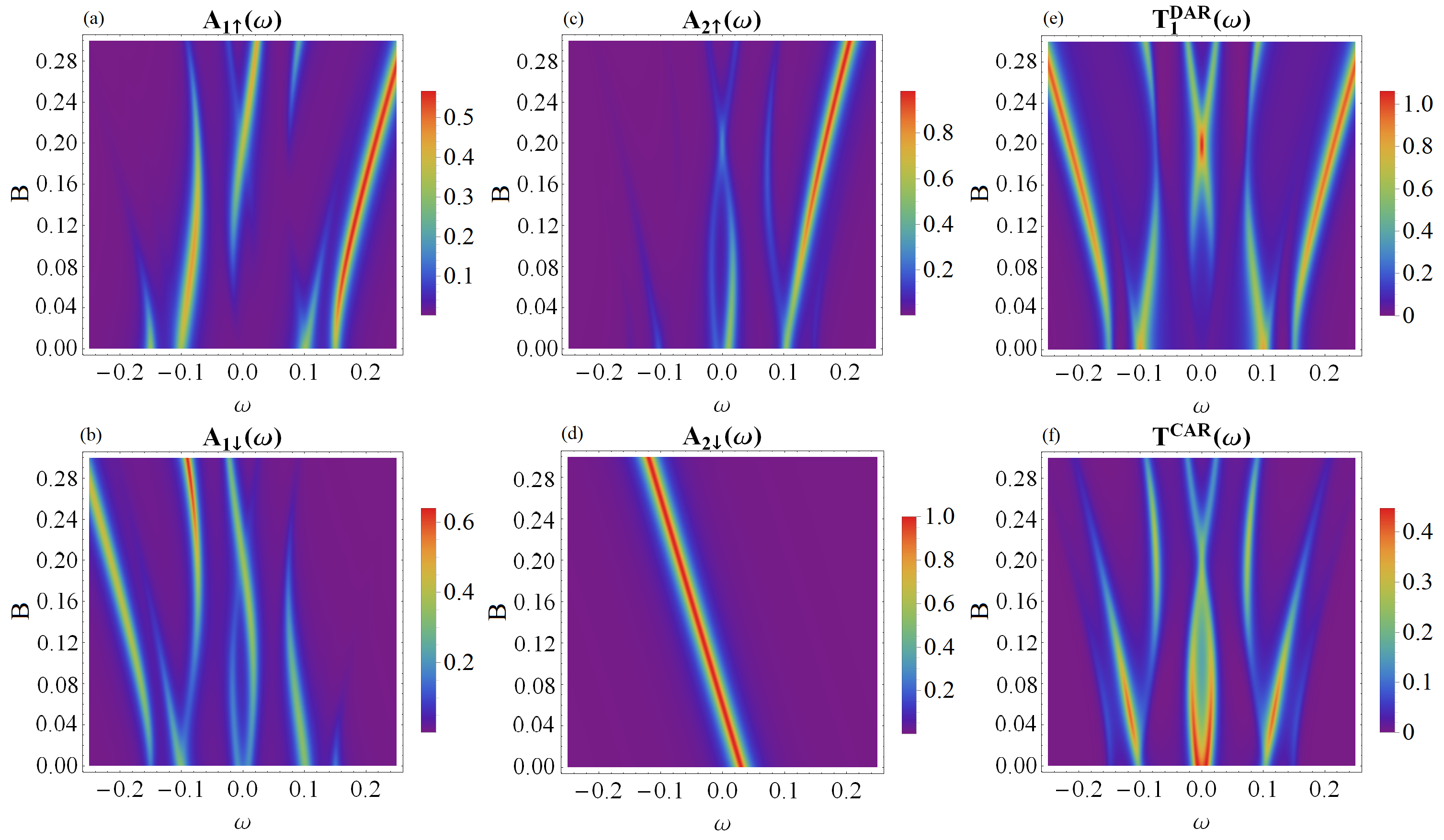}
	\caption{
 The spin-dependent spectral functions of QD$_{1}$ (a and b panels) and QD$_2$ (c and d panels) and the transmittance of the DAR (e) and CAR (f) tunneling channels with respect to the Zeeman field $B$. The results are obtained for the model parameters: $\epsilon_1=0$, $\epsilon_2=0.03$, $\en_M=0.05$, $t_{1}=t_{2}=0.05$, $t'_{1}=t'_{2}=0$, and $\theta=\pi$.}
	\label{FigZeeman}
\end{figure}
\end{widetext}


\section{
Influence of Zeeman field}
\label{sec:Zeeman}

Topologically nontrivial superconductivity of semiconducting nanowires arises upon
combining the Rashba interaction with the superconducting proximity effect in the presence of sufficiently strong external magnetic field (usually on the order of
$1$~T). The same criterion is necessary for the minimally short-length topological nanowires \cite{Kouewenhoven-2023,Bordin2024Feb}, where the Majorana modes might overlap between themselves. For this reason, it is important to check the influence of magnetic field on the quasiparticle spectra of our hybrid structure (Fig.~\ref{scheme}), taking into account the Zeeman splitting of the
quantum dot levels, $B=\varepsilon_{i\downarrow}-\varepsilon_{i\uparrow}$. 

Similar issue has been previously addressed by Prada et al. \cite{Prada-2017} for a single quantum dot attached to the short topological wire, predicting either bowtie-like (for  $t_{1}' \ll \epsilon_{M}, t_{1}$) or diamond-like (for $\epsilon_{M} \ll t_{1}, t_{1}'$) superstructure of the quasi-Majorana modes, respectively. For brevity, we focus here on the linear configuration ($t_{1}'=0=t_{2}'$) and analyze evolution of the spin-resolved quasiparticles of the quantum dots with respect to the Zeeman field $B$ (see Fig.~\ref{FigZeeman}). In close analogy to the results reported
in Refs \cite{Prada-2017,Gorski-2020} we obtain the bowtie feature in the spin-$\uparrow$ spectral functions of both quantum dots. Furthermore, we notice this bowtie-like shape appearing in the spin-$\downarrow$ spectral function of QD$_{1}$, which is indirectly driven by the on-dot pairing between the opposite spins.
Such bowtie-structure is in turn evident, both in the direct and crossed Andreev transmittance (see right h.s. panels in Fig.~\ref{FigZeeman}),
thus it would be detectable experimentally.
As far as the higher-energy (trivial) quasiparticles are concerned,
they reveal rather complex variation against the Zeeman field, depending on
various spin arrangements of the quantum dots and additionally being affected by the on-dot pairing at QD$_{1}$ (discussed at length in the main part of our manuscript).

\acknowledgements{
This research project has been supported by National Science Centre (Poland) through the grant 
no.\ 2022/04/Y/ST3/00061.
}

\section*{Data availability}
The datasets generated and analyzed during the current study are available
from the repository \cite{data_2024} and/or  upon request from the corresponding author.

\bibliography{mybib2}

\end{document}